\documentclass[sigconf,authorversion]{aamas}  % do not change this line!

\usepackage{booktabs}

\setcopyright{ifaamas}  % do not change this line!
\acmDOI{doi}  % do not change this line!
\acmISBN{}  % do not change this line!
\acmConference[AAMAS'18]{Proc.\@ of the 17th International Conference on Autonomous Agents and Multiagent Systems (AAMAS 2018), M.~Dastani, G.~Sukthankar, E.~Andre, S.~Koenig (eds.)}{July 2018}{Stockholm, Sweden}  % do not change this line!
\acmYear{2018}  % do not change this line!
\copyrightyear{2018}  % do not change this line!
\acmPrice{}  % do not change this line!

\begin{document}

\title{Local Wealth Redistribution Promotes Cooperation in Multiagent Systems} 

\author{Fl\'{a}vio L. Pinheiro}
\orcid{0000-0002-0561-9641}
\affiliation{
  \institution{Collective Learning Group, The MIT Media Lab}
  \city{Massachusetts Institute of Technology, 22 Ames Street, Cambridge} 
  \state{Massachusetts} 
  \postcode{02139}
}
\email{flaviopp@mit.edu}

\author{Fernando P. Santos}
\affiliation{%
  \institution{INESC-ID and Instituto Superior T\'{e}cnico, \\Universidade de Lisboa}
  \state{2744-016 Porto Salvo, Portugal}
}
\email{fernando.pedro@tecnico.ulisboa.pt}

\begin{abstract}  % put your abstract here!
Designing mechanisms that leverage cooperation between agents has been a long-lasting goal in Multiagent Systems. 
The task is especially challenging when agents are selfish, lack common goals and face social dilemmas, \textit{i.e.}, situations in which individual interest conflicts with social welfare. 
Past works explored mechanisms that explain cooperation in biological and social systems, providing important clues for the aim of designing cooperative artificial societies. 
In particular, several works show that cooperation is able to emerge when specific network structures underlie agents' interactions. 
Notwithstanding, social dilemmas in which defection is highly tempting still pose challenges concerning the effective sustainability of cooperation. 
Here we propose a new redistribution mechanism that can be applied in structured populations of agents. 
Importantly, we show that, when implemented locally (\textit{i.e.}, agents share a fraction of their wealth surplus with their nearest neighbors), redistribution excels in promoting cooperation under regimes where, before, only defection prevailed.
\end{abstract}



  % put your semicolon-separated keywords here!

\maketitle

%%%%%%%%%%%%%%%%%%%%%%%%%%%%%%%%%%%%%%%%%%%%%%%%%%%%%%%%%%%%%%%%%%%%%%%%%%%%%%%%%%%%%%%%%%%%%%%%%%%%%%%%%
%% start of main body of paper

\section{Introduction}

Explaining cooperation among selfish and unrelated individuals has been a central topic in evolutionary biology and social sciences \cite{nowak2012evolving}. 
Simultaneously, the challenge of designing cooperative Multiagent Systems (MAS) has been a long standing goal of researchers in artificial intelligence (AI) \cite{genesereth1984cooperation,jennings1998roadmap}.
More than thirty years ago it was already clear that \textit{''Intelligent agents will inevitably need to interact flexibly with other entities. 
The existence of conflicting goals will need to be handled by these automated agents, just as it is routinely handled by humans.``} \cite{genesereth1984cooperation}.

In Cooperative multiagent interactions, agents need to collaborate towards common goals, which introduces challenges associated with coordination, communication and teamwork modeling \cite{panait2005cooperative,jennings1998roadmap}.
Self-interested interactions, in contrast, require the design of indirect incentive schemes that motivate selfish agents to cooperate in a sustainable way \cite{ephrati1996deriving,jennings1998roadmap}.
Cooperation is often framed as an altruistic act that requires an agent to pay a cost ($c$) in order to generate a benefit ($b$) to another.
Refusing to incur in such a cost is associated with an act of defection and results in no benefits generated.
Whenever the benefit exceeds the cost ($b>c$) and plays occur simultaneously, agents face the Prisoner's Dilemma, a decision-making challenge that embodies a fundamental social dilemma within MAS \cite{macy2002learning}: rational agents pursuing their self-interests are expected to defect, while the optimal collective outcome requires cooperation.
If defection is the likely decision of rational agents, however, how can we justify the ubiquity of cooperation in the real world?
Evolutionary biology has pursued this fundamental question by searching for additional evolutionary mechanisms that might help to explain the emergence of cooperative behavior \cite{nowak2012evolving,nowak2006five}.
Some of these mechanisms allowed to develop solutions that found applications in computer science, such as informing about ways of incentivizing cooperation in \textit{p2p} networks \cite{golle2001incentives,feldman2005overcoming}, wireless sensor networks \cite{akyildiz2002wireless}, robotics \cite{waibel2011quantitative} or resource allocation and distributed work systems \cite{seuken2010accounting} -- to name a few.

Network reciprocity is one of the most popular mechanisms to explain the evolution of cooperation in social and biological systems \cite{nowak1992evolutionary,santos2005scale,ohtsuki2006simple,pinheiro2012local,pinheiro2016linking,pinheiro2017intermediate}.
In this context, populations are structured and interactions among agents are constrained.
These constraints are often modelled by means of a complex network of interactions.
Applications of this mechanism have been explored in the design of MAS that reach high levels of cooperation \cite{hofmann2011evolution,peleteiro2014exploring,ranjbar2014theory,airiau2014emergence}. 
Despite these advances, cooperation on structured populations is still hard to achieve when considering social dilemmas with high levels of temptation to defect. Additional complementary mechanisms are required.

Here we consider that agents contribute a percentage of their surplus (defined below), which is later divided among a Beneficiary Set of other agents. %(see Figure \ref{fig:Figure1}).
In this context, we aim at answering the following questions:

\begin{itemize}
    \item Does redistribution of wealth promote the evolution of cooperation?
    \item How should Beneficiary Sets be selected?
    \item What are the potential disadvantages of such a mechanism?
\end{itemize}

Using methods from Evolutionary Game Theory (EGT) \cite{sigmund2010calculus} and resorting to computer simulations, we explore how wealth redistribution impacts the evolution of cooperation on a population of agents without memory (\textit{i.e.} unable to recall past interactions) and rationally bounded (\textit{i.e.} lacking full information on payoff structure of the game they are engaging). We assume that agents resort to social learning through peer imitation, which proves to be a predominant adaptation scheme employed by humans \cite{rendell2010copy}. Also, we consider that strategies are binary -- Cooperate and Defect -- opting to focus our attention on the complexity provided by 1) heterogeneous populations, 2) the redistribution mechanism and 3) the self-organizing process of agents when adapting over time. The role of larger strategy spaces (such as in \cite{peleteiro2014exploring,ranjbar2014theory,santos2017structural}) lies outside the scope of the present work.

With redistribution, we show that cooperation emerges in a parameter region where previously it was absent.
Moreover, we show that the optimal choice of redistributing groups consists in picking the nearest neighbors (local redistribution). 
This result fits with a local and polycentric view of incentive mechanisms \cite{ostrom2015governing,vasconcelos2015cooperation} in MAS, which may not only be easier to implement but, as we show, establish an optimal scale of interaction in terms of eliciting cooperation.

\section{Related Work}
The problem of Cooperation is a broad and intrinsically multidisciplinary topic, which has been part of the MAS research agenda for a long time \cite{jennings1998roadmap,genesereth1984cooperation}.
In the realm of evolutionary biology, several mechanisms were proposed to explain the evolution of cooperation \cite{nowak2006five}. 
Kin selection \cite{hamilton1964genetical}, direct reciprocity \cite{trivers1971evolution}, indirect reciprocity \cite{nowak2005evolution,santos2018social} and network reciprocity \cite{santos2005scale,ohtsuki2006simple} constitute some of the most important mechanisms proposed. 
Remarkably, these mechanisms have been applied in AI in order to design MAS in which cooperation emerges. 
For example, \citeauthor{waibel2011quantitative} associated kin selection with evolutionary robotics \cite{waibel2011quantitative}; \citeauthor{griffiths2008tags} employed indirect reciprocity to promote cooperation in \textit{p2p} networks while \citeauthor{ho2012towards} investigated the social norms that, through a system of reputations and indirect reciprocity, promote cooperation in crowdsourcing markets \cite{griffiths2008tags,ho2012towards}. 
Similarly, \citeauthor{peleteiro2014exploring} combined indirect reciprocity with complex networks to design a MAS where, again, cooperation is able to emerge \cite{peleteiro2014exploring}. 
On top of that, \citeauthor{han2016emergence} applied EGT -- as performed in our study -- in order to investigate the role of punishment and commitments in multiagent cooperation, both in pairwise \cite{han2016emergence} and group interactions \cite{han2017centralized}. Regarding alternative agent-oriented approaches to sustain cooperation in MAS, we shall underline the role of electronic institutions \cite{esteva2004ameli,arcos2005engineering} whereby agents' actions are explicitly constrained so that desirable collective behaviors can be engineered.

The role of population structure and network reciprocity is, in this context, a prolific area of research. 
In \cite{pinheiro2012local} it was shown that complex networks are able to fundamentally change the dilemma at stake, depending on the particular topology considered \cite{pinheiro2012local,ichinose2017mutation}; \citeauthor{ranjbar2014theory} applied tools from control theory in order to study the role of complex networks on the evolution of cooperation \cite{ranjbar2014theory}. 
Importantly, the role of dynamic networks -- \textit{i.e.}, agents are able to rewire their links -- was also shown to significantly improve the levels of cooperation, especially in networks with a high average degree of connectivity \cite{santos2006cooperation,pinheiro2016linking}. 
A survey on the topic of complex networks and the emergence of cooperation in MAS can be accessed in \cite{hofmann2011evolution}.

Previous works found that cooperation in structured population substantially decreases when the temptation to defect increases (see Model for a proper definition of Temptation).
Thereby, here we contribute with an additional mechanism of cooperation on structured populations.
We consider a mechanism of redistribution, inspired in the wealth redistribution mechanisms that prevail in modern economic/political systems, mainly through taxation. 
We are particularly interested in understanding how to sample redistribution groups in an effective way. 
In this context, we shall underline the works of \citeauthor{salazar2011emerging} and \citeauthor{burguillo2009memetic}, in which a system of taxes and coalitions was shown to promote cooperation on complex networks \cite{salazar2011emerging} and regular grids  \cite{burguillo2009memetic}. While \cite{salazar2011emerging} and \cite{burguillo2009memetic} do an excellent job showing how coalitions -- leaded by a single agent -- emerge, here we consider a simpler/decentralized model (e.g. no leaders are considered and taxes are redistributed rather than centralized in a single entity) and focus our analysis on showing that local redistribution sets are optimal. Our approach does not require additional means of reciprocity, memory, leadership, punishment or knowledge about features of the network. We cover a wide range of dilemma strengths and explicitly show when the local redistribution promotes cooperation by itself. Notwithstanding, the analysis performed in \cite{salazar2011emerging} and \cite{burguillo2009memetic} surely provides important insights to address in future works, on how to explicitly model the adherence to beneficiary sets and guarantee their stability.
Also, while here we assume an egalitarian redistribution over each individual in the Beneficiary Set, we shall note that different redistribution heuristics may imply different levels of allocation fairness \cite{pitt2012self}. In this context, a recent work introduces the concept of Distributed Distributed Justice \cite{kurka2016distributed} and shows that local interactions may provide a reliable basis to build trust and reputation between agents, which can be used to regulate, in a decentralized way, the levels of justice in agents' actions. This way, it is rewarding to note that local interactions not only constitute an optimal scale to form cooperative Beneficiary Sets (as we show, see below), but also provide the convenient interaction environment to allow justice in contributions to be sustained.

\section{Model}
\subsection{Three Stage Redistribution Game}
Here we propose a sequential game dynamics made of three stages.
Focusing on an arbitrary agent $i$, these stages can be described as follows:
\begin{enumerate}
	\item Agent $i$ participates in a one-shot game (here a Prisoner's Dilemma) with all his/her neighbors $j$. From each interaction $j$, he/she obtains a payoff $\pi_{i,j}$. After all interactions, agent $i$ accumulates a total payoff $\Pi_i=\sum_{j}\pi_{i,j}$;
	\item Next, agent $i$ contributes a fraction $\alpha$ of his/her payoff surplus ($\Pi_i - \theta$) to be redistributed. The group that benefits from agent $i$ contribution is called Beneficiary Set $i$ ($B_i$).
	\item Finally, agent $i$ receives his/her share from each Beneficiary Set that he/she is part of.
\end{enumerate}

We refer to $\alpha$ as the level of taxation, as it defines the fraction of the surplus that agents contribute, while $\theta$ is the threshold level of payoff that defines the surplus. 
By definition, agents with negative payoff cannot contribute (\textit{i.e.}, $\theta > 0$); they might, however, receive benefits from the Beneficiary Sets.
Each agent $i$ contributes only to one Beneficiary Set $B_i$ from which they cannot be part of, that is, agents do not receive from the Beneficiary Set they contribute to.
A central question of this work is how to select $B_i$ for each $i$. 
As we show, this decision has a profound and non-trivial impact on the overall cooperation levels in the system. 

\subsection{The Prisoner's Dilemma Game}
\begin{figure}[!t]
    \centering
    \includegraphics[width=.8\columnwidth]{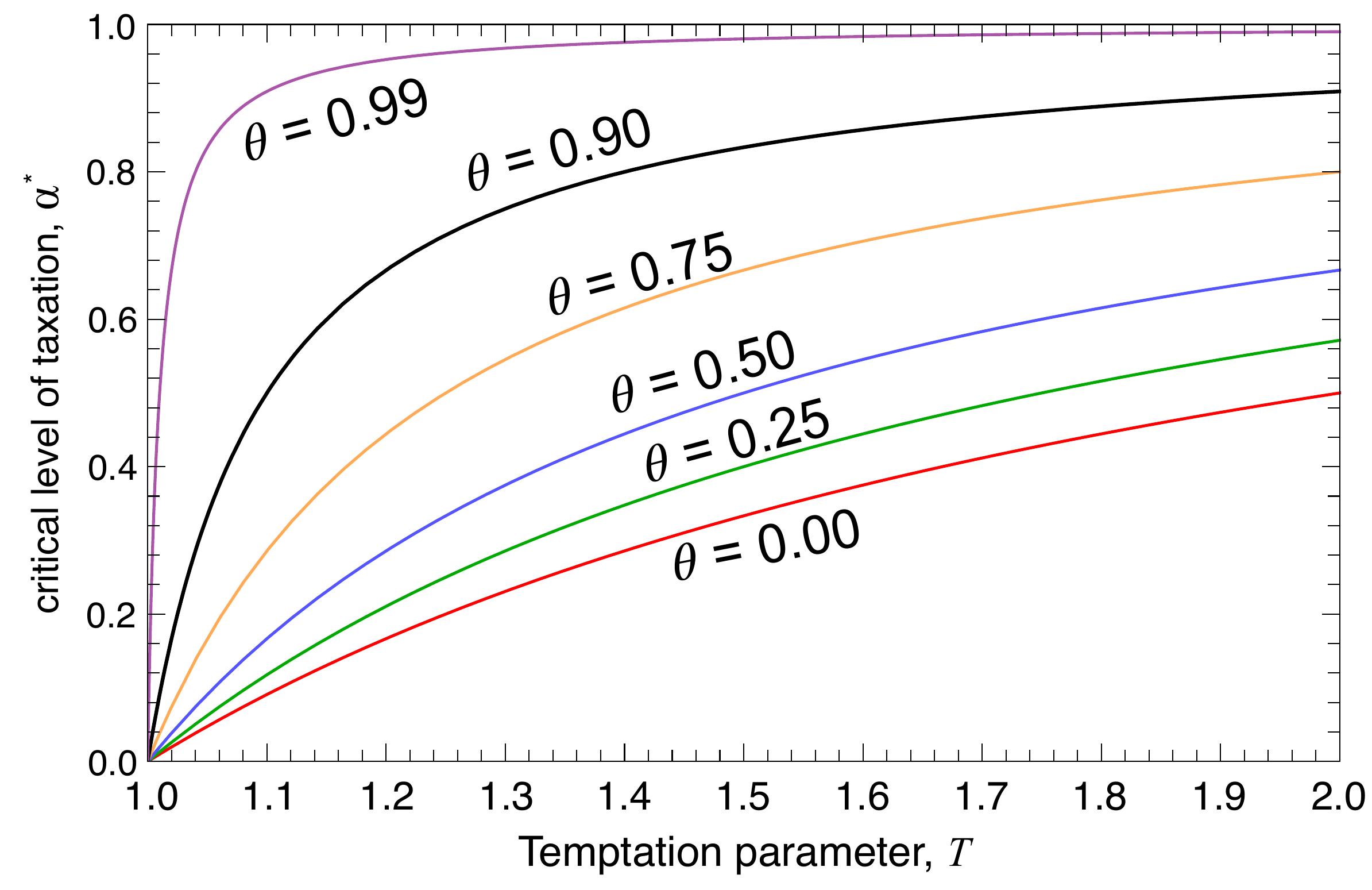}
    \caption{Solutions for the two-person game with wealth redistribution. Each curve indicates the critical taxation levels ($\alpha^\ast$) above which the nature of the social dilemma changes, for different payoff thresholds ($\theta$) and as a function of the Temptation parameter ($T$).}
    \label{fig:Figure2}
\end{figure}

In general, all the possible outcomes of a two-strategy two-player game, in which two agents engage in a one-shot interaction that requires them to decide -- independently and simultaneously -- whether they wish to Cooperate (C) or to Defect (D), can be summarized in a payoff matrix, such as
\bgroup
\def\arraystretch{1.5}
\begin{center}
    \begin{tabular}{ c | c c }
            & C     & D \\
        \hline		
        C   & $R$       & $S$ \\
        D   & $T$       & $P$ \\
    \end{tabular}
\end{center}
which reads as the payoff obtained by playing the row strategy when facing an opponent with the column strategy. Here, $R$ represents the Reward payoff for mutual cooperation and $P$ the Punishment for mutual defection. 
When one of the individuals Defects and the other Cooperates, the first receives the Temptation payoff ($T$) while the second obtains the Sucker's payoff ($S$).
In this manuscript we consider that agents interact according to the Prisoner's Dilemma (PD).
Agents are said to face a PD whenever the relationship between the payoffs is such that $T > R > P > S$ \cite{sigmund2010calculus}.  
In such a scenario, rational agents seeking to optimize their self-returns are expected to always Defect. 
However, since the best aggregated outcome would have both players cooperating ($2R > 2P$), agents are said to face a social dilemma: optimizing self-returns clashes with optimizing the social outcome. In this sense, mutual cooperation is Pareto Optimal and contributes to increase both average payoff (over mutual defection) and egalitarian social welfare (over unilateral cooperation) \cite{endriss2003welfare}.
It is noteworthy to mention that other situations -- with different optimal rational responses -- arise when the parameters take a different relationship \cite{macy2002learning}: the Stag Hunt game when $R > T > P > S$; the Snowdrift Game when $T > R > S > P$; the Harmony Game when $R > T > S> P$; or the Deadlock Game when $T > P > R > S$, to name a few. Notwithstanding, the PD is by far the most popular metaphor of social dilemmas \cite{sigmund2010calculus} and the one that presents the biggest challenge for cooperation to emerge. 
For these reasons, PD shall be the main focus of study in this manuscript. 
We further simplify the parameter space by considering that $ R= 1$, $P = 0$, $S = 1 - T$ and $1 < T \leq 2$ with the game being fully determined by the Temptation value (T).
In that sense, higher temptation creates more stringent conditions for the emergence of cooperation.

\subsection{Prisoner's Dilemma with Wealth Redistribution}
\begin{figure}[!t]
    \centering
    \includegraphics[width=0.8\columnwidth]{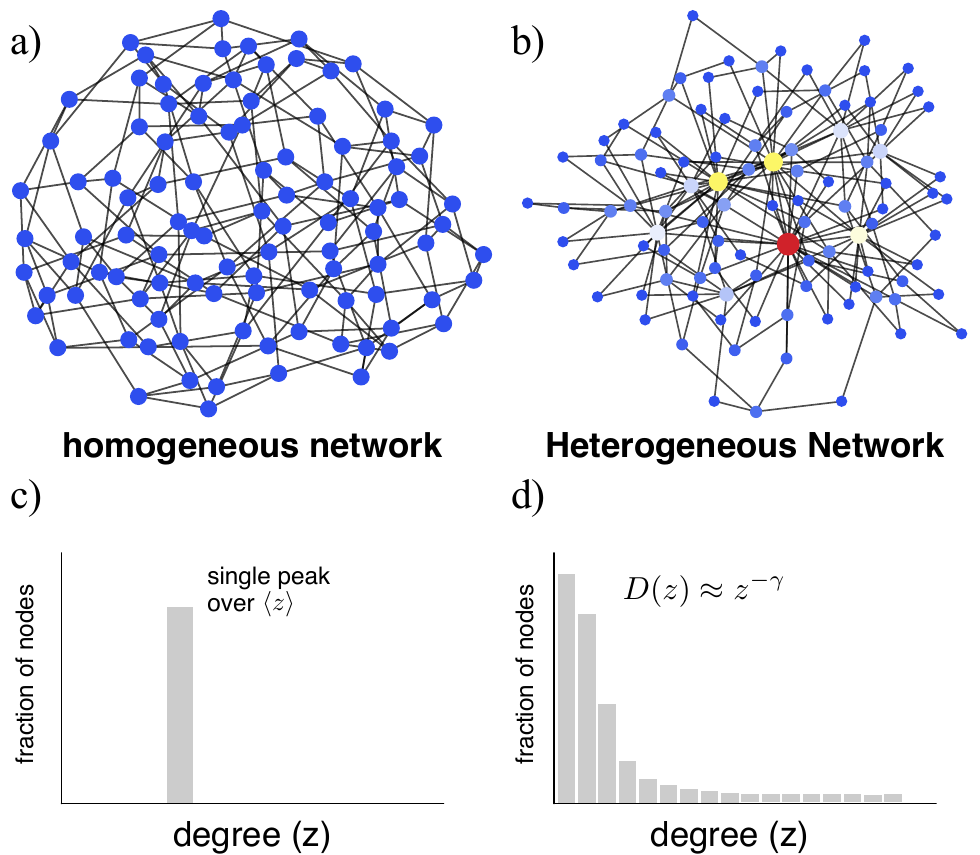}
    \caption{Graphical depiction of the specific structures used in this work.
        a) Homogeneous Networks correspond to a structure in which all nodes have the same degree.
        b) Heterogeneous Networks are characterized by a high variance among the degree of nodes. The color of each node indicates its degree: blue tones represent lower degree and red tones higher degree.
        Panel c) and d) show, respectively, the degree distributions of the Homogeneous and Heterogeneous networks under analysis. In particular, we use scale-free networks as representatives of heterogeneous structures; these have a degree distribution that decays as a power law.
    }
    \label{fig:Figure3}
\end{figure}

As an introductory example, let us start by analyzing the particular case of two interacting agents ($i$ and $j$) in a one-shot event. 
In this case, the Beneficiary Sets of each agent ($B_i$ and $B_j$) are composed only by the opponent. 
Wealth/payoff redistribution can thus be analyzed by considering a slightly modified payoff matrix, that takes into account the second and third stages. 
The resulting payoff matrix becomes
\bgroup
\def\arraystretch{1.5}
\begin{center}
    \begin{tabular}{ c | c c }
            & C     & D \\
        \hline		
        C   & $1$   & $1-T + \alpha(T-\theta) $ \\
        D   & $T - \alpha(T-\theta) $   & $0$ \\
    \end{tabular}
\end{center}
where $\theta$ is the payoff threshold and $\alpha$ is the level of taxation. 
The rationale to arrive at this payoff structure is the following: whenever both players choose to act the same way the payoff remains the same as their contributions (from taxes) and benefits (from receiving the contributions of their opponent) cancel out. 
A Defector playing against a Cooperator sees his payoff of $T$ subtracted by an amount $\alpha(T-\theta)$ while not receiving any benefit, since the Cooperator has negative payoff and does not contribute. 
Likewise, the Cooperator is exempt from contributing but receives an additional contribution of $\alpha(T-\theta)$, which represents the amount taxed to the Defector.

\begin{figure}[!t]
    \centering
    \includegraphics[width=0.8\columnwidth]{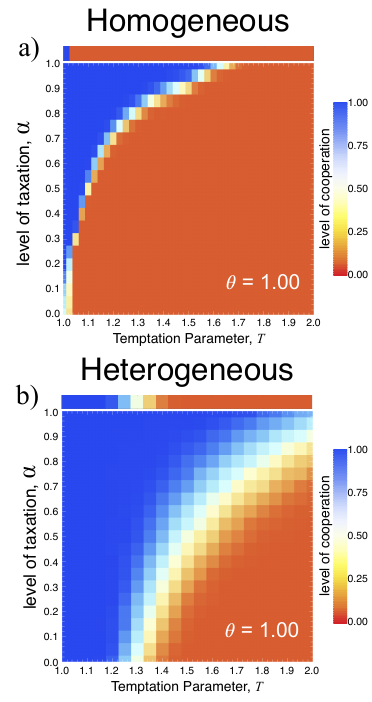}
    \caption{ Level of Cooperation on Homogeneous Random Networks (a) and Heterogeneous (Scale-free) Networks (b). Each plot shows the level of cooperation under a different combination of taxation level, $\alpha$, and Temptation, $T$. In all cases the fitness threshold is fixed at $\theta = R = 1.0$. Blue indicates regions where Cooperation dominates, Red delimits regions dominated by Defectors. Top bars above each panel indicate the level of cooperation in the absence of wealth redistribution, as a function of the Temptation payoff parameter. The level of cooperation is computed by estimating the expected fraction of cooperators when the population reaches a stationary state. To that end we run $10^4$ independent simulations that start with $50\%$ cooperators and $50\%$ defectors. Population size of $Z = 10^3$ and intensity of selection $\beta = 1.0$.}
    \label{fig:Figure4}
\end{figure}

To inspect whether wealth redistribution changes the nature of the social dilemma (\emph{i.e.} from a Prisoner's Dilemma to another type of game) we have to inspect whether there is a difference in the relationship between the payoffs $R$ and $T$ or $P$ and $S$.
This sums up to solving a single inequality,
\begin{equation}
        T - \alpha(T-\theta) < 1
\end{equation}
which results in the critical values of $\alpha$,
\begin{equation}
\label{eq:gtsol}
     \alpha^\ast > \frac{T-1}{(T-\theta)}
\end{equation}
Hence, depending on the choice of $\theta$ and for a given $T$, $\alpha^\ast$ is the minimum level of taxation required to observe a change in the nature of the game faced by agents. 
It is straightforward to notice that the nature of the game changes from a Prisoner's Dilemma to an Harmony Game as the relationship moves from $T>R>P>S$ to $R>T>S>P$. 
Figure~\ref{fig:Figure2} shows $\alpha^\ast$ for different values of  $T$ and  $\theta$. 
Clearly, in well-mixed populations and under the simple scenario of a MAS composed by two agents, the redistribution mechanism has the simple effect of reshaping the payoff matrix, trivially changing the nature of the dilemma. 
Such a trivial conclusion cannot be drawn with large populations playing on networks, where we will show that different ways of assigning the Beneficiary Sets have a profound impact on the ensuing levels of cooperation. 

\begin{figure}[!t]
    \centering
    \includegraphics[width=0.8\columnwidth]{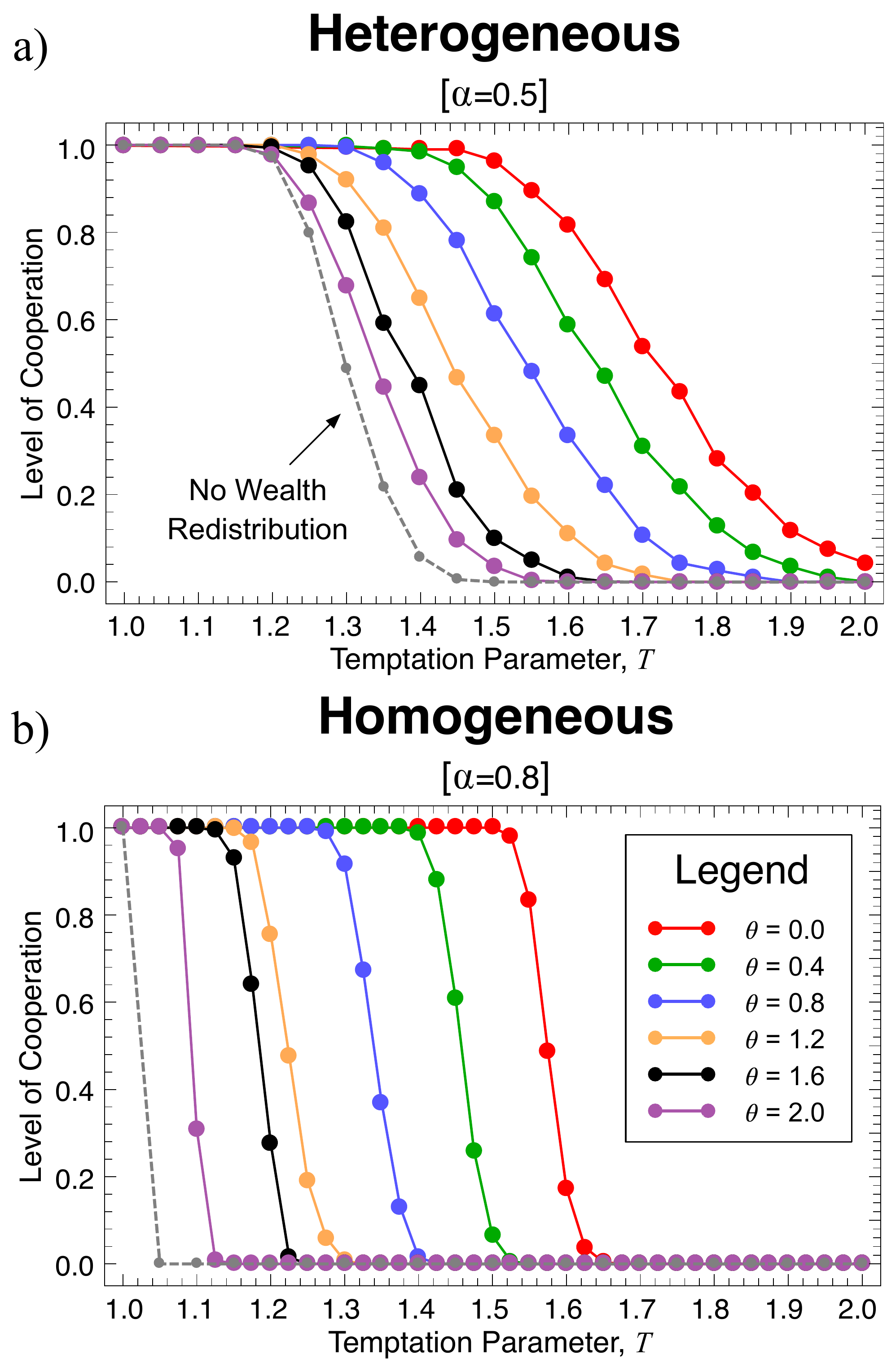}
    \caption{Level of cooperation on Heterogeneous (a) and Homogeneous (b) populations for different values of the payoff threshold ($\theta$) as a function of the Temptation payoff parameter ($T$). Gray Dashed line shows the results obtained in the absence of a wealth redistribution scheme. Population size of $Z = 10^3$ and intensity of selection $\beta = 1.0$.
    }
    \label{fig:Figure5}
\end{figure}

\subsection{Structured Populations}
Let us consider a population of $Z$ agents in which agents correspond to the nodes/vertices of a complex network, while links dictate who interacts with whom. 
The structure reflects the existence of constraints that limit interactions between agents. 
These constraints can arise from spatial or communication limitations. 

The number of interactions that each agent $i$ participates in defines his/her degree $z_i$. 
The distribution of degrees, $D(z)$, describes the fraction of agents that has degree $z$. 
In this work we consider two structures: Homogeneous Random Graphs \cite{santos2005epidemic,santos2017structural} and Scale-Free Barab\'{a}si Networks \cite{albert2002statistical}.

Homogeneous Random Graphs are generated by successively randomizing the ends of pairs of links from an initially regular graph (\textit{e.g.} Lattice or Ring). 
The resulting structure has a random interaction structure but all nodes in the network have the same degree. 
Figure~\ref{fig:Figure3}a) depicts graphically an example of such structures and Figure~\ref{fig:Figure3}c) the corresponding Degree distribution.

Scale-free networks are generated by an algorithm of growth and preferential attachment \cite{albert2002statistical}. This algorithm is as follows: 1) start from three fully connected nodes; 2) add, sequentially, each of the $Z-m$ remaining nodes; 3) each time a new node is added, it connects to $m$ pre-existing nodes, selecting preferentially nodes with higher degree. Here we have used $m = 3$ The resulting network is characterized by a heterogeneous degree distribution (one which decays as a power law), in which the majority of the nodes have few connections while a few have many. Figure~\ref{fig:Figure3}b) shows a graphical example of such structure and Figure~\ref{fig:Figure3}d) the degree distribution.

In the following we explore the case of  networks with $Z=10^3$ nodes and average degree of $\langle z \rangle = \sum_z z D(z)  = 4$.
During the simulations we make use of $20$ independently generated networks of each type.

\subsection{Games on Networks}

We study the expected level of cooperation attained by the population.
We estimate this quantity through computer simulations. 
The level of cooperation corresponds to the expected fraction of cooperators in a population that evolved after $2.5\times 10^6$ iterations. 
We estimate this quantity by averaging the observed fraction of cooperators at the final of each simulation, over $10^4$ independent simulations.

Each simulation starts from a population with an equal composition of Cooperators and Defectors, which are randomly placed along the nodes of the network. 
In between each update round, each agent $i$ plays once with all his/her $z_i$ nearest neighbors (\textit{i.e.}, agents they are directly connected with). 
The accumulated payoff over all interactions an agent $i$ participates can be computed as
\begin{equation}
    \Pi_i = n^C_i T - \sigma_i (1-T) (n^D_i + n^C_i)
\end{equation}
where $n_i^D$ ($n_i^C$) is the number of neighbors of $i$ that Defect (Cooperate) and $\sigma_i$ is equal to 1 if $i$ is a Cooperator and 0 otherwise. 
From the accumulated payoff, agents contribute to a pool a fraction $\alpha$ of the surplus $\Pi-\theta$. 
The fitness $f_i$ of an agent $i$ results from subtracting from his/her accumulated payoff his/her contributions plus the share he/she obtains from each of the Beneficiary Sets $j$ he/she participates in. We shall underline that, while $T$ is the same for all agents (that is, the dilemma is the same for everyone in the population), heterogeneous populations introduce an additional complexity layer by implying that different agents may vary in the maximum values of accumulated payoff that they are able to earn.
This can be formalized as
\begin{equation}
	f_i = (1-\alpha)(\Pi_i-\theta)+\sum_j^Z  \delta_{i,j}\frac{\alpha(\Pi_j-\theta)}{|B_j|}
\end{equation}
where $\delta_{i,j}$ is equal to one if $i$ is part of the Beneficiary Set towards which $j$ contributes and zero otherwise, while $|B_j|$ denotes the size of set $B_j$.
\begin{figure}[!t]
    \centering
    \includegraphics[width=0.8\columnwidth]{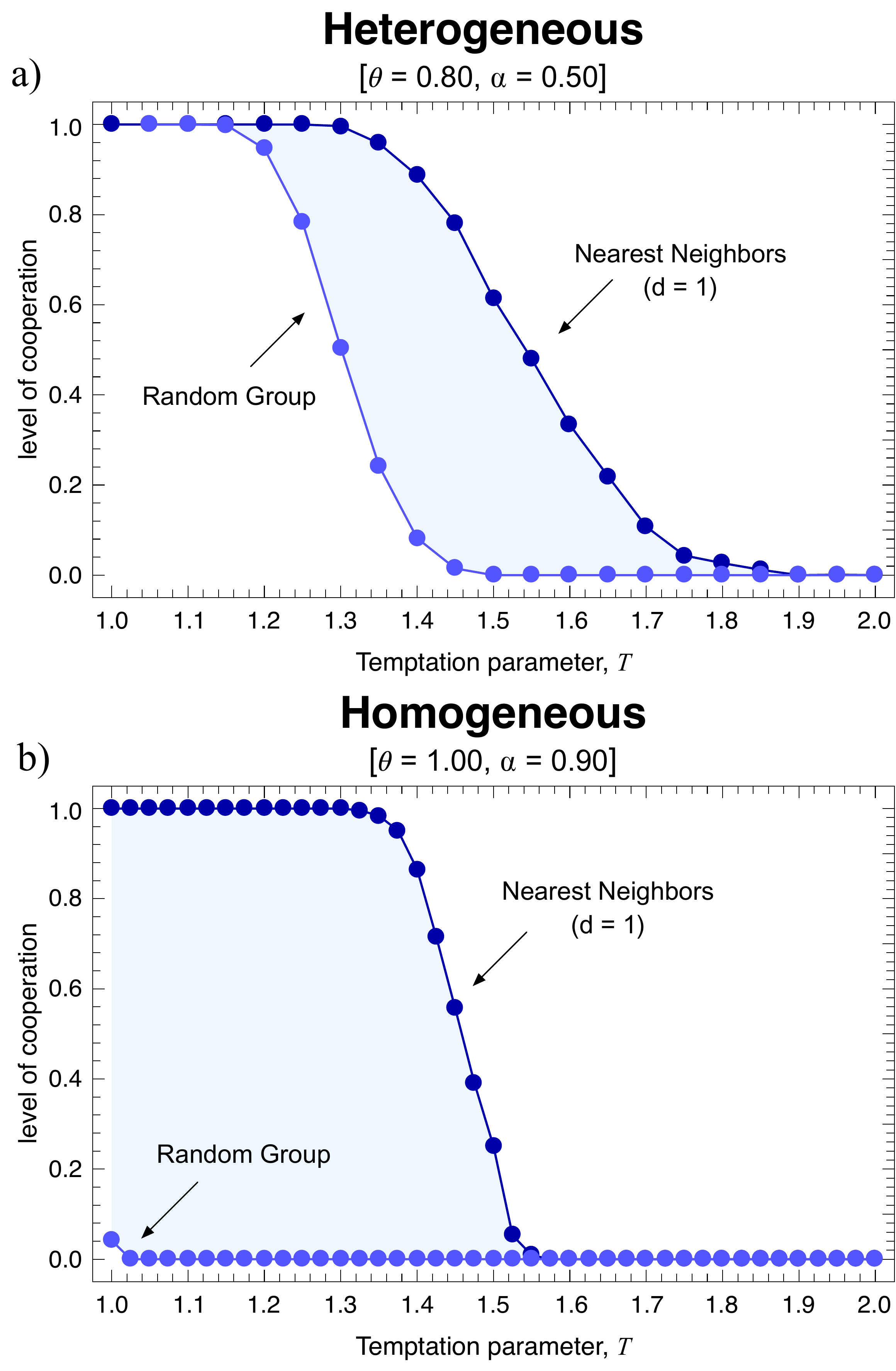}
    \caption{Comparison between the effects of assigning the nearest neighbors of an agent $i$ to the corresponding Beneficiary Set $B_i$ (dark blue line) and when agents are assigned at random to $B_i$ (light blue), on the level of cooperation in the domain of the Temptation payoff parameter, T. Panel a) shows the results on Heterogeneous populations and panel b) the impact on Homogeneous populations. Population size of $Z = 10^3$ and intensity of selection $\beta = 1.0$.
    }
    \label{fig:Figure6}
\end{figure}

Evolution in the frequency of strategies adopted in the population happens through a process of imitation or social learning. 
At each iteration a random agent, say $i$, compares his fitness with the fitness of a neighbor, say $j$. Depending on the fitness difference, $i$ adopts the strategy of $j$ with probability 
\begin{equation}
    p = \frac{1}{1+Exp(-\beta(f_j - f_i))}
\end{equation}
The meaning of this sigmoid function can be understood as follows: if $j$ is performing much better than $i$, then $i$ updates his/her strategy, adopting the strategy of $j$. 
Conversely, if $j$ is performing much worse, $i$ does not update the strategy. 
The parameter $\beta$, often called the intensity of selection and akin to a learning rate, dictates how sharp is the transition between these two regimes, as $f_j - f_i$ approaches zero. 
Large $\beta$ means that individuals act in a more deterministic way, updating strategies at the minimum difference; small $\beta$ means that individuals are prone to make imitation mistakes.
\section{Results}
\subsection{Wealth Redistribution and the Level of Cooperation in Structured Populations}
In this section we start by analyzing the scenario in which the Beneficiary Set of each agent $i$ corresponds to his/her nearest neighbors.
Hence, the size of the Beneficiary set of $i$ is $|B_i| = z_i$.
These are also the agents from whom he/she interacts with and obtains a payoff from.
Figure~\ref{fig:Figure4} shows the achieved levels of cooperation  when the payoff threshold is set to $\theta = R = 1.0$, as a function of the Temptation payoff ($T$) and the level of taxation ($\alpha$). 
Figure~\ref{fig:Figure4}a shows the results on Homogeneous networks, and Figure~\ref{fig:Figure4}b on Heterogeneous.  
We find that, for a fixed payoff threshold ($\theta$), increasing the level of taxation results in an increase in the levels of cooperation. 
This effect diminishes with an increase in the Temptation  ($T$). 
That is, when increasing $T$ the minimum value of $\alpha$ necessary to promote cooperation increases as well. 
The same behavior is observed in both structures. 
However, there is a larger degree of cooperation on Heterogeneous networks, where there is always a level of taxation for a given Temptation that guarantees a 100$\%$ level of cooperation.
Hence, in order for cooperation to be evolutionary viable on homogeneous networks, more stringent conditions are necessary, \textit{e.g.} higher tax levels.

Figure~\ref{fig:Figure5} shows how the level of cooperation depends on variations of the fitness threshold ($0 \leq \theta \leq 2.0$, in intervals of $0.4$) while keeping a fixed level of taxation ($\alpha = 0.5$) under different levels of the Temptation payoff ($T$).
Figure~\ref{fig:Figure5}a shows the results obtained for Heterogeneous networks and panel b) the results on Homogeneous structures. 
For a constant level of taxation, $\alpha$, decreasing the payoff threshold, $\theta$, increases the range of Temptation, $T$, under which cooperation can possibly evolve. 
This is the case in both types of structures. 
However, once again, the effect is more limited in homogeneous populations.

Both Figure~\ref{fig:Figure4} and \ref{fig:Figure5} highlight the positive impact of a local wealth redistribution mechanism in the enhancement of cooperation. 
It also puts in evidence that the success of such mechanism depends on the volume of payoff that is redistributed. 
Ultimately, this can be done by either increasing the level of taxation, $\alpha$ or decreasing the payoff threshold, $\theta$, that defines the taxable payoff.
\begin{figure}[!t]
    \centering
    \includegraphics[width=0.8\columnwidth]{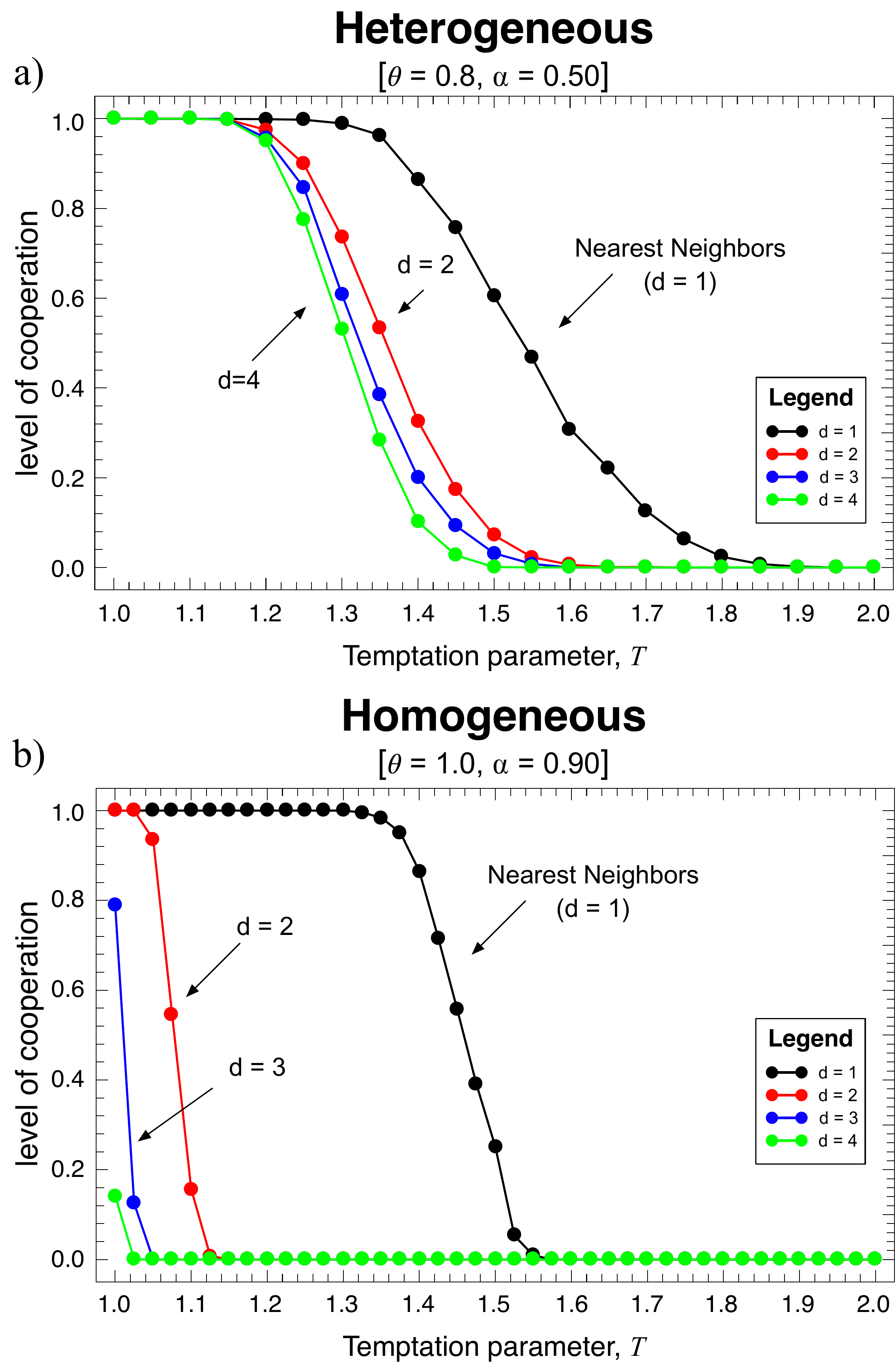}
    \caption{Panel a) compares how extending beneficiary sets, from the nearest neighbors ($d=1$) to nodes at a distance up to $d = 4$ links away, impacts the level of cooperation on Heterogeneous networks. Panel b) shows how extended beneficiary sets impact the level of cooperation on Homogeneous networks. In both cases extending the set of beneficiaries has a negative a negative impact in the levels of cooperation. Population size of $Z = 10^3$ and intensity of selection $\beta = 1.0$.
    }
    \label{fig:Figure7}
\end{figure}
\subsection{Randomized Beneficiary Set}
Next we explore to which extent the results obtained depend on the way agents are being assigned to each Beneficiary Set.
To that end, we compare two cases: i) nearest set assignment --  the Beneficiary Set of each agent corresponds to her/his nearest neighbors, as above; and ii) random set assignment -- agents are assigned at random to each Beneficiary Set. The number of agents assigned to each set is equal to the degree of the contributing agent, in both cases, which guarantees that the collected payoffs from each agent are distributed among the same number of individuals in both i) and ii). 

Figure~\ref{fig:Figure6}a and b show the results obtained, respectively, on Heterogeneous  and Homogeneous networks. 
We consider $\theta = 0.5$, $\alpha = 0.9$ and explore the domain $1.0 \leq T \leq 2$. 
Dark blue curves show the results obtained under the nearest set assignment and light blue curves the results obtained under a random set assignment. 
The results show that the ability of a wealth redistribution mechanism lies in the redistribution of the taxed payoff among the agents that are spatially related.
A random assignment of agents drastically decreases the levels of cooperation obtained in both networks.
But to which extent do the Beneficiary Sets need to be constrained spatially?

\begin{figure}[!t]
    \centering
    \includegraphics[width=0.8\columnwidth]{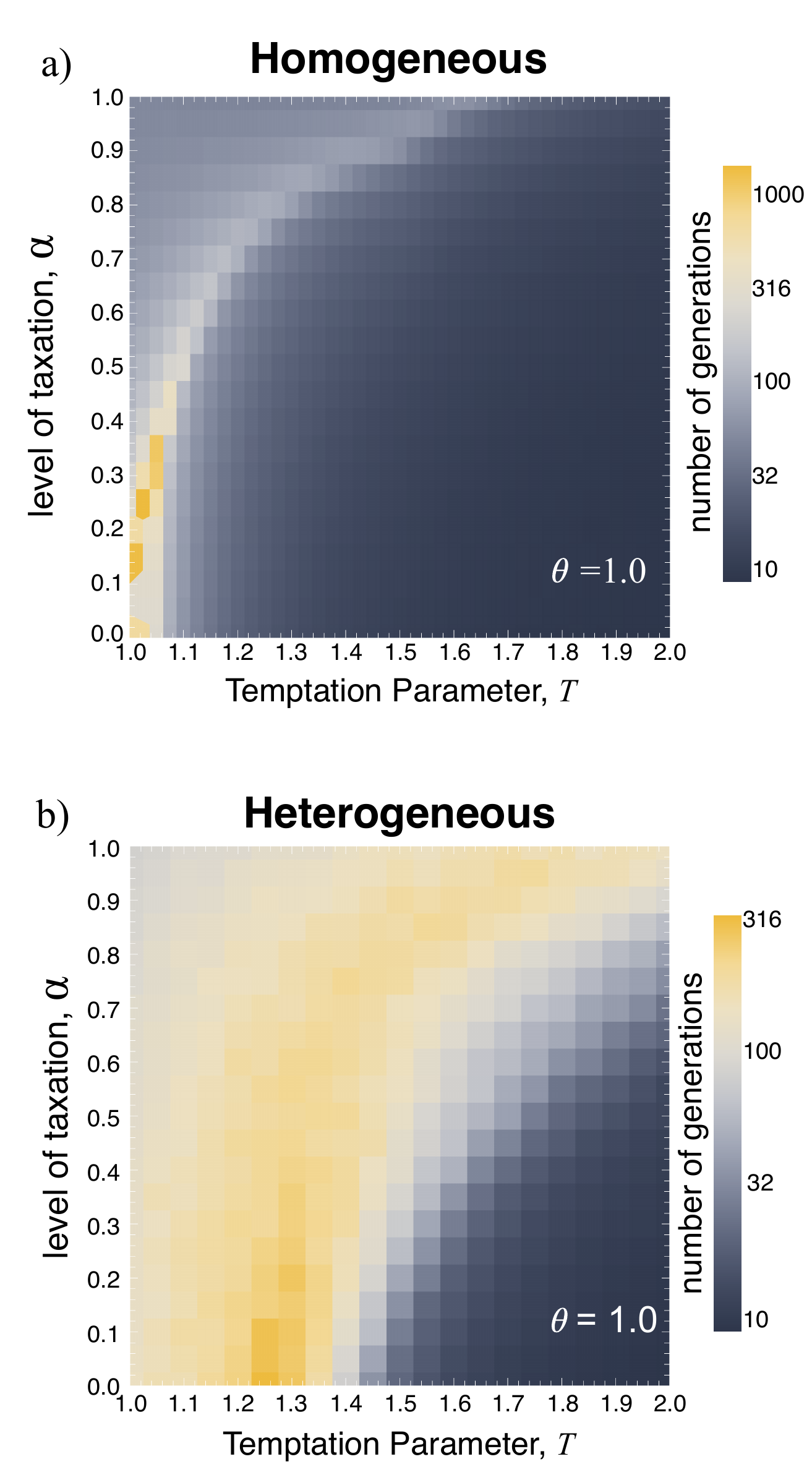}
    \caption{Panel a) shows the fixation times (in generations) on homogeneous networks. Panel b) shows the fixation times (in generations) in heterogeneous networks. A generation corresponds to $Z$ iteration steps and the fixation times indicate the expected time that the population takes to arrive to a state dominated by Cooperators or Defectors when starting from a state with equal abundance of both strategies. Population size of $Z = 10^3$ and intensity of selection $\beta = 1.0$.
    }
    \label{fig:Figure8}
\end{figure}

\subsection{Extended Beneficiary Set}
To answer the previous question, we explore the case in which all nodes (up to a distance of $d$ links) are assigned to the Beneficiary Set of a focal agent $i$; when $d=1$ the previous results are thereby obtained.

Figure~\ref{fig:Figure7}a and b show the results up to $d=4$ on Heterogeneous and Homogeneous networks respectively.
In both cases, we see that an expansion in the size of the Beneficiary Set leads to a decrease in the levels of cooperation.
This result further reinforces the conclusion that wealth redistribution is only efficient when agents return, in form of taxes, a share of the accumulated payoffs to the agents they have engaged with.
We shall underline that here both distance and size of $B_i$ play a role on the obtained results, while in the previous section the size of $B_i$ was kept constant for each $i$ across the different treatments, thus disambiguating the effect of $B_i$ size and distance on the resulting cooperation levels.

\subsection{What is the cost of wealth redistribution?}
Figure~\ref{fig:Figure8}a and b shows the fixation times of populations when $\theta = 1.0$ along the domain bounded by $0.0 \leq \alpha \leq 1.0$ and $1.0 \leq T \leq 2.0$. 
The fixation times correspond to the expected number of generations (i.e., sets of $Z$ potential imitation steps) for the population to reach a state in which only one strategy is present in the population.
These plots map directly into Figure~\ref{fig:Figure4}a and b, allowing to compare the relative fixation times of regions with high/low levels of cooperation. 

We observe that the evolution of cooperation is associated with an increase in the fixation times.
This increase can be in some situations an order of magnitude higher. 
The regions that exhibit larger fixation times lie in the critical boundary that divides areas of defectors and cooperators dominance (Figure~\ref{fig:Figure4}).
Hence, promoting cooperation by redistributing wealth also requires a longer waiting time for the population to reach a state of full cooperation.
However, setting higher taxation values than the bare minimum necessary for the emergence of cooperation allows populations to reach fixation quicker.

\subsection{\label{sec:ContributionLevels}Multiple Contribution Brackets}

%\begin{figure}[!t]
%    \centering
%    \includegraphics[width=0.7\columnwidth]{Figure9.pdf}
%    \caption{Level of cooperation with multiple taxation brackets. Panel a) shows results for $\theta = 0.8$ and panel b) shows results for $\theta = 2.0$. Variation in the number of taxation brackets has a marginal impact in the overall levels of cooperation. Population size of $Z = 10^3$ and intensity of selection $\beta = 1.0$.
%    }
%    \label{fig:Figure9}
%\end{figure}

In the real world, taxes are unlikely to be defined by a single threshold ($\theta$) that separates agents who contribute from those that do not.
In reality taxes are progressive, in the sense that taxation levels ($\alpha$) increase with increasing level of income (in this case accumulated payoff). In this section we implement a similar approach and inspect the impact of increasing the number of taxation brackets.

Let us consider that, instead of a single threshold we now have $B$ taxation brackets divided by $B-1$ threshold levels.
For each bracket we define $\alpha_b$ as the effective tax and $\theta_b$ as the bottom threshold of bracket $b$, where $b\in \{0,1,2,...,B-1,B\}$.

By definition $B = 0$ corresponds to the case in which no taxes are collected, and the redistribution of wealth is absent. Moreover, $B = 1$, implies the existence of a single bracket were all individuals would contribute, a case that we do not explore in this manuscript. $B = 2$, corresponds to the case in which there are two brackets, which is the scenario that we have explored until now. 

We consider the case in which taxation increases linearly with increasing brackets.
Let us define $\theta_b = b\theta/B$. Individuals in bracket $b$ have their payoff surplus taxed by $\alpha_b = (b-1)\alpha/B$ when their accumulated payoff falls into $\theta_b < \Pi \leq \theta_{b+1} $ for $b < B-1$.
For $b = B$ the tax level is $\alpha_b = \alpha$ and affects all individuals with $\Pi > \theta $.

As an example, for $B = 3$ each bracket would be characterized by the following tax levels
\begin{itemize}
    \item[$b = 0$)] $\alpha_b = 0$ for all individuals with $\Pi \leq \theta/3 $;
    \item[$b = 1$)] $\alpha_b = \alpha/3$ for all individuals with $ \theta/3 < \Pi \leq 2\theta/3 $;
    \item[$b = 2$)] $\alpha_b = 2\alpha/3$ for all individuals with $ 2\theta/3 < \Pi \leq \theta $;
    \item[$b = 3$)] $\alpha_b = \alpha$ for all individuals with $\Pi > \theta $.
\end{itemize}
In this way we use $\theta$ and $\alpha$ as the upper level bound and only parameters in this condition.

We find that variations in the number of taxation brackets ($B$=3,4,5) have only a marginal impact in the overall levels of cooperation observed when compared with the scenarios studied so far ($B$=2). 

\subsection{\label{sec:Inequality}Wealth Inequality}

\begin{figure}[!t]
    \centering
    \includegraphics[width=0.8\columnwidth]{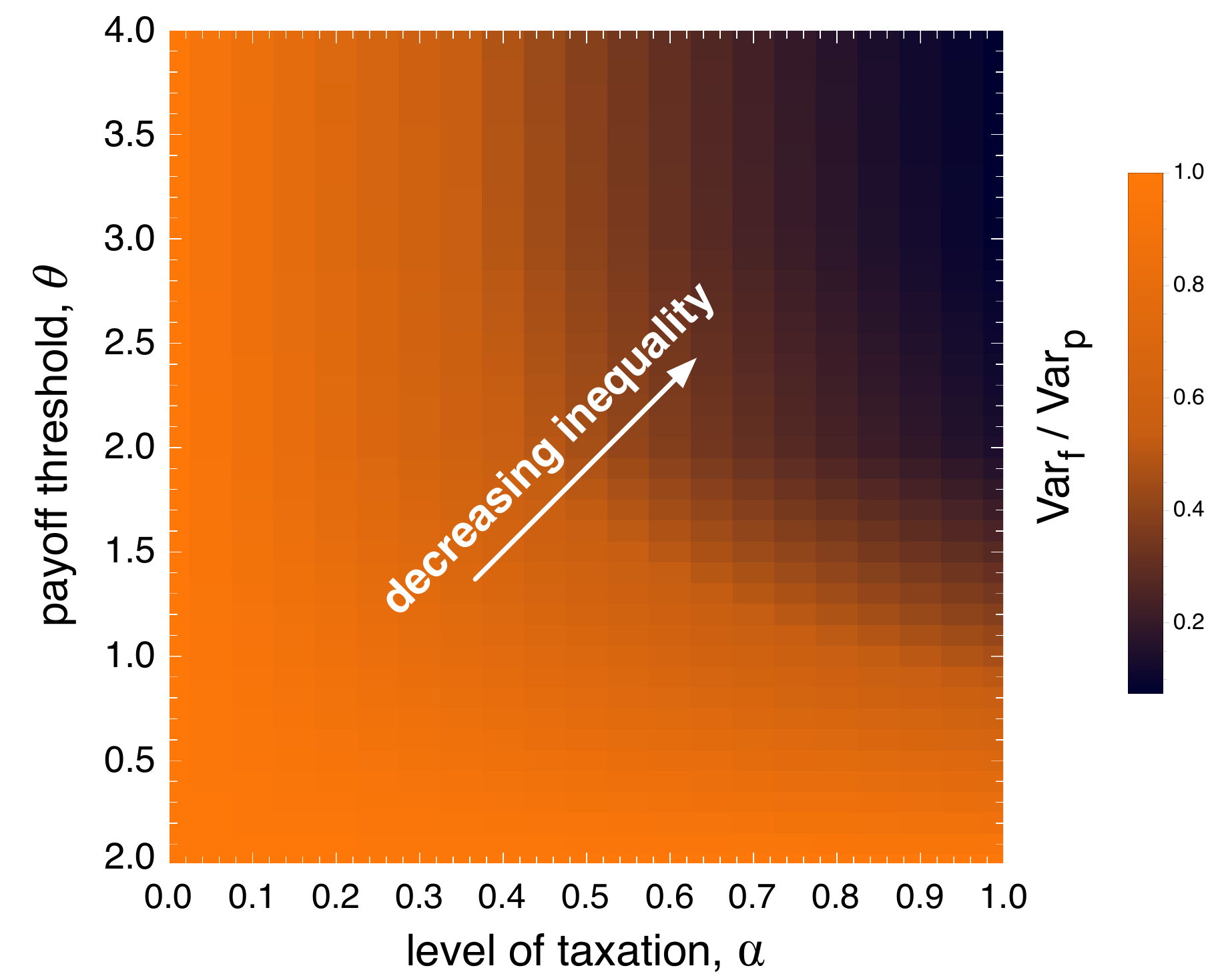}
    \caption{Relative wealth inequality after the redistribution step, in a heterogeneous population dominated by cooperators and for different combinations of taxation level ($\alpha$) and threshold ($\theta$). We quantify the relative wealth inequality after the redistribution step as the ratio between the variance of the fitness distribution ($Var_{f}$, i.e. variance in gains across the population after redistribution) and the variance of the accumulated payoff distribution ($Var_{p}$, i.e. variance in gains before redistribution). Population size of $Z = 10^3$ and intensity of selection $\beta = 1.0$.
    }
    \label{fig:Figure10}
\end{figure}

Finally, we discuss the effect of wealth redistribution on fitness inequality. 
First, it is important to highlight that the observed levels of inequality depend, by default, on the distribution of strategies and network degree. In homogeneous structures, if every agent adopts the same strategy -- either Defectors or Cooperators -- everyone obtains the same fitness. In heterogeneous structures, a Cooperation dominance scenario bounds the feasible equality levels, given the degree distribution of the population. In fact, some agents engage in more interactions than others and Beneficiary Sets have different sizes, depending on the particular connectivity of agents. We shall focus on this scenario.

We compare the variance of fitness (\textit{i.e.} gains after the redistribution step) and the variance of accumulated payoff (\textit{i.e.} gains before the redistribution step) in order to quantify the relative inequality after we apply the proposed redistribution mechanism. In particular, we use the ratio between the variance of fitness and the variance of accumulated payoff as a metric of resulting wealth inequality.

Figure~\ref{fig:Figure10} shows how higher levels of $\theta$ and $\alpha$ reduce the resulting inequality. In fact, while increasing payoff threshold limits taxation to the richer agents, increasing level of taxation increases the flow of fitness from rich agents to their Beneficiary Sets. In the most strict case -- high $\theta$ and $\alpha$ --  the variance of the fitness distribution is reduced to as low as $7\%$ of the accumulated payoff distribution.

\section{\label{sec:Conclusion}Conclusion}
To sum up, we show that wealth redistribution embodies an effective mechanism that significantly helps cooperation to evolve. It works by fundamentally changing the nature of the dilemma at stake: by appropriately choosing the level of taxation ($\alpha$) and payoff threshold ($\theta$) it is possible to shift from a Defector dominance to a Cooperator dominance dynamics. Moreover, we find that in Heterogeneous populations allow us to ease the redistribution mechanism -- that is, imposing lower taxation rates and/or lower taxable surplus values when compared with Homogeneously structured populations.

Additionally, we show, for the first time, that different assignments of Beneficiary Sets significantly impact the ensuing levels of cooperation. Local Beneficiary Sets, where agents receive the contributions from their direct neighbors, constitute a judicious choice when compared with Beneficiary Sets that are formed by 1) agents randomly picked from the population or 2) by including agents at higher distances. Naturally, a Local wealth redistribution scheme may not only prove optimal in terms of achieved cooperation levels, but also reveal much simpler to implement, by exempting the need of central redistribution entities and by minimizing the number of peers that agents need to interact with. We shall highlight, however, that promoting cooperation through a wealth redistribution mechanism bears longer fixation times, in terms of the number of iterations required to achieve overall cooperation. 

Here we assume that the redistribution mechanism is externally imposed. Agents are not able to opt out from the taxation scheme. Given that this mechanism increases the overall cooperation and average payoff in the system, an argument for its acceptance - by rational agents - can be formulated based on the infamous veil of ignorance proposed by John Rawls \cite{rawls2009theory}: Agents should decide the kind of society they would like to live in without knowing their social position. Agents would, this way, prefer a cooperative society where redistribution exists, provided that here average payoff is maximized. Notwithstanding, future research shall analyze the role of more complex strategies that give opportunity of agents to voluntarily engage (or not) in the proposed redistribution scheme. Alongside, effective mechanisms that discourage the second order free riding problem (i.e., free riding by not contributing to the redistribution pot, while expecting others to do so) shall be examined. Future works shall also evaluate whether alternative taxation schemes are prone to be more efficient than the one proposed here. In all these cases, an evolutionary game theoretic framework -- such as the one here developed -- constitutes a promising toolkit to employ.

%na versao a submeter tem que se retirar
\section{Acknowledgments}
The authors acknowledge the useful discussions with Francisco C. Santos, Jorge M. Pacheco and Aamena Alshamsi.
F.L.P. is thankful to the Media Lab Consortium for financial support.
F.P.S. acknowledges the financial support of Funda\c{c}\~{a}o para a Ci\^{e}ncia e Tecnologia (FCT) through PhD scholarship SFRH/BD/94736/2013, multi-annual funding of INESC-ID (UID/CEC/50021/2013) and grants PTDC/EEI-SII/5081/2014, PTDC/MAT/STA/3358/2014.

%%%%%%%%%%%%%%%%%%%%%%%%%%%%%%%%%%%%%%%%%%%%%%%%%%%%%%%%%%%%%%%%%%%%%%%%%%%%%%%%%%%%%%%%%%%%%%%%%%%%%%%%%
%% bibliography: see CFP for number of permitted pages

\bibliographystyle{ACM-Reference-Format}  % do not change this line!
\bibliography{References}  % put name of your .bib file here

\end{document}